# A Novel FPGA-Based High Throughput Accelerator For Binary Search Trees

Oyku Melikoglu, Oguz Ergin
Department of Computer Engineering
TOBB University of Economics and Technology
Ankara, Turkey
{omelikoglu, oergin} @etu.edu.tr

Behzad Salami, Julian Pavon, Osman Unsal, Adrian Cristal
Computer Architecture for Parallel Paradigms Group
Barcelona Supercomputing Center
Barcelona, Spain
{behzad.salami, julian.pavon, osman.unsal, adrian.cristal}
@bsc.es

*Abstract*—**This paper presents a deeply pipelined and massively parallel Binary Search Tree (BST) accelerator for Field Programmable Gate Arrays (FPGAs). Our design relies on the extremely parallel on-chip memory, or Block RAMs (BRAMs) architecture of FPGAs. To achieve significant throughput for the search operation on BST, we present several novel mechanisms including tree duplication as well as horizontal, duplicated, and hybrid (horizontal-vertical) tree partitioning. Also, we present efficient techniques to decrease the stalling rates that can occur during the parallel tree search. By combining these techniques and implementations on Xilinx Virtex-7 VC709 platform, we achieve up to 8X throughput improvement gain in comparison to the baseline implementation, i.e., a fully-pipelined FPGA-based accelerator.**

## I. INTRODUCTION

Binary Search Tree (BST) is a traditional and fundamental data structure. It is widely-used in the structure of many state-of-the-art applications such as database, machine learning, file systems, among others. BST is based on storing data items in a sorted format within a tree structure and thus, results in reducing the search time complexity to log (n), where n is the number of items in the tree. Lookup, Insert, and Delete are the most common operations on BST. Among them, Lookup operation is used to find a certain data item, i.e., key in the tree. In this paper, we aim to accelerate the Lookup operation of BST using a hardware-level optimization specified for Field Programmable Gate Arrays (FPGAs). FPGAs are widely-used computing devices in the state-of-the-art High-Performance Computing (HPC) systems, thanks to their massively parallel architecture and the capability of stream-fashion data execution models. Also, their energy dissipation is significantly less than other more flexible computing devices such as CPUs and Graphics Processing Units (GPUs). Furthermore, with the rise of High-Level Synthesis (HLS) tools to facilitate the FPGA application development, exploiting FPGAs in data centers is becoming mainstream. It is expected that FPGAs will be in 30% of data centers by 2020 [15].

Effectively exploiting the inherent features of FPGAs mentioned above can lead to highly-optimized designs such as, usage for database systems [7] , neural networks [13,14] , hash join [8,9] , and sort [10, 12] algorithms. However, although, there are hardware-based BST accelerators [1-6], they are not fully optimized to take the advantage of inherent parallelism and pipelining capability of FPGA architecture. To alleviate this issue, this paper aims to build a relatively higher-throughput FPGA-based accelerator for BST search algorithm. In our accelerator, the parallel structure of FPGA components especially on-chip memories, or Block RAMs (BRAMs), play the key role. Parallel accesses to different BRAMs and performing the required comparison operations in a pipelined fashion are the core optimization techniques exploited in our accelerator. In particular, we partition the tree within different BRAMs and build the required hardware to traverse the tree in a parallel and pipelined manner. Toward this goal, depending on the trade-off, we present tree duplication as well as horizontal, duplicated, and hybrid (horizontal-vertical) partitioning methods. We propose tree duplication as an orthogonal optimization technique, which aims to achieve relatively higher throughput by duplicating the tree on BRAMs. In addition, horizontal tree partitioning, i.e., locating each level of the tree in different memories, can take advantage of the deep pipelining. Also, with vertical tree partitioning the inherent parallelism capability of FPGAs is exploited to achieve further higher throughput. In the hybrid partitioning approach, which is developed to take advantage of both pipelining and parallelism search on the tree without any duplication, we are also increasing the throughput by adding buffers to decrease the stalling rates that can occur during a search.

In short, the main contributions of this paper are listed as below:

- We exploit the inherently parallel BRAMs structure of FPGAs to accelerate the BST search. We demonstrate the effectiveness of this approach by evaluating several methods, e.g., tree duplication and partitioning with different trade-offs on the latency, throughput, and resource utilizations.

- To maximize the bandwidth utilization rate of BRAM, we exploit buffers to decrease the cycle stalling rate that can occur during the search of the tree in a parallel and pipelined fashion. Toward this goal, we present direct and queue mapping techniques.

We implement the proposed BST tree on a VC709 platform, with Virtex7 FPGA and evaluate it with different types of key sets. The most optimized technique that we evaluated achieves up to 8X higher throughput in comparison to the state-of-the-art fully-pipelined FPGA-based accelerator. The system design parameters such as buffer sizes, number of trees or vertical partitions are fully reconfigurable at compile time. Thus, throughput is flexible and can even be further increased trading off throughput against power, frequency and memory.

The rest of the paper is organized as follows. Section 2, elaborates the proposed accelerator for the BST search, i.e., discussing the proposed techniques in detail to achieve a significant throughput. The experimental results and consequent achievements are discussed in Section 3. Section 4, briefly reviews the related work. Finally, Section 5 summarizes and concludes the paper.

## II. HARDWARE BASED ACCELERATED SEARCH

We store the data which is composed of 32 Bit Key and 32 Bit Value pairs as a binary tree. Our experiments are based on a complete binary tree as the throughput will not change when the type of tree changes during a stream of infinite keys. The binary tree data is stored inside BRAMs. BRAMs can be synchronously written and read. The BRAMs can be used separately or can be configured to be combined with each other. Groups of BRAMs will be referred as BRAM Partitions.

The keys to be searched are fetched as chunks. We are selecting the size of the chunks equal to the maximum number of keys that can be searched parallel in a single cycle.

The objective is to increase the throughput of the key search, therefore we present three different methods, horizontal, duplicated and hybrid partitioning, which are specialized for FPGAs.

### A. Horizontal Partitioning

In this approach we store every level of the tree in a different BRAM Partition to be accessed in parallel thus increasing throughput. For instance, the root is stored in the first BRAM Partition and a node with a height of four is stored in the fifth partition. Fig. 1 shows the relations between levels of the binary tree and the BRAM Partitions.

In commercial FPGAs, BRAMs are mainly dual-port. We exploit this capability to achieve a 2X higher throughput than single-port configuration. Since BRAM Partitions are a larger, combined version of the selected type of BRAM, their port attributes are the same. This approach results in a search of again at most one key per port at the same time inside the same

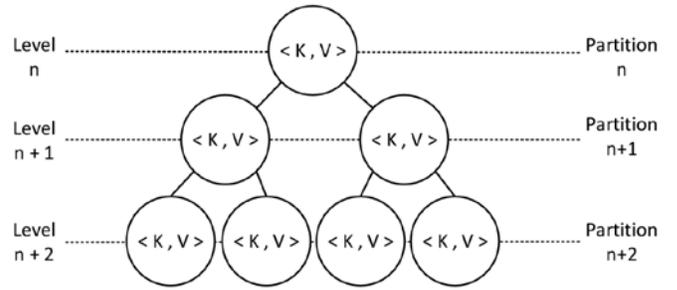

Figure 1. BRAM Partitions corresponding the levels

BRAM Partition. However, instead of waiting for the keys to be found while stalling new keys to commence with the search process, the method can be pipelined. By using different BRAM Partition groups, all of the levels of the tree can be reached at the same time. While one key is searched inside the first partition, another can be searched in the last one. This leads to an increase for the number of keys that can be searched at the same time and it enables a search of h (height of the tree) + 1 keys at the same time, which are searched in different levels of the tree. By taking advantage of this property, we implement a pipelined way of searching keys.

Starting from the root of the tree, we compare $Key_s$ (The key that is searched) with the root node, if $Key_s$ is bigger than $Key_n$ (The key of the corresponding node) the search continues to the right child of the node, if $Key_s$ is smaller than $Key_n$, the search continues to the left child of the node. Because the children of the nodes are in a different BRAM Partition than the corresponding node, in the next clock cycle $Key_s$ is searched in the next BRAM Partition. Since $Key_s$ is no longer being searched in the previous partition, instead of letting the partition stay idle $Key_{s+1}$ is started to be searched. In every cycle a new key's search starts. This results in no idle BRAM Partitions after h clock cycles if all of the keys are not found before the leaf nodes of the tree. With the pipelining approach, the throughput is two keys per clock cycle. This pipelining approach is shown in Fig. 2.

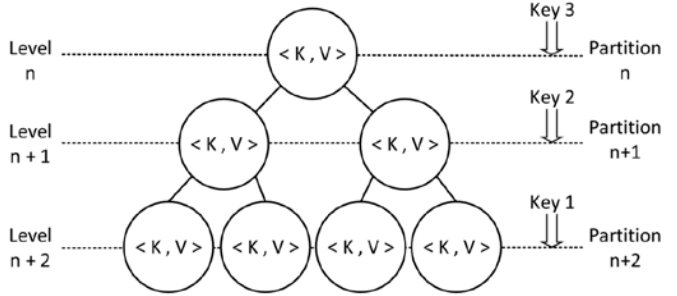

Figure 2. Pipelining between Partitions

## B. Duplicated Horizontal Partitioning

With the purpose of increasing the throughput, we implement the duplication method, which means storing multiple copies of the tree and thus, allows parallel search. We do the BRAM Partitioning the same way as the prior method. Every level of the tree is stored in a different partition and every tree is stored within different partitions, that is no two trees contain a common partition. With this approach we increase the throughput from two keys per cycle to "number of tree replica x2" keys per cycle. However, the data needs to be duplicated for additional trees resulting in a need for more storage space.

## C. Hybrid Partitioning

### 1) Addition of Registers

We divide the tree into two different layers. First layer consists of registers which contain the first levels of the tree. The rest of the tree is stored in the second layer which consists of BRAM partitions. Each dual-port BRAM Partition is inherently limited in its capability (two per cycle) to quickly store the maximum number of keys before the search process can start, while the registers do not have this limitation. This is the motivation for the Hybrid Partitioning idea. Furthermore, the hybrid design is further optimized by assigning the first layer to registers, since the root needs to be traversed in every search for accessing other layers consisting of BRAM Partitions. Unlike BRAMs, registers are capable of being accessed more flexibly, that is since registers do not have ports, they can be read simultaneously. Therefore if the first levels of the tree is put inside the registers instead of BRAMs, more keys can start the search at the same time. However if the keys are not found and they need to be compared with the keys which are not stored in the registers, but stored in the BRAMs, a bottleneck is going to occur. This bottleneck is due to the fact that, the number of keys that can be searched simultaneously will be again decreased to one key per port in the worst case. However in the best case scenario this number will be $2^x$ where x corresponds to the number of levels in the register layer since all of the keys will be directed to a different BRAM Partition.

Another advantage of using registers as a storage for first levels is using the BRAMs more efficiently. That is, if instead of using register layer, all nodes were put in partition layers even though there is only one node in a level, a BRAM Partition is spared for that level. However by storing the levels which does not have sufficient nodes to fill a partition in registers we fully utilize BRAMs.

### 2) Horizontal and Vertical Partitioning

In order to remove the bottleneck, and continue using the BRAMs, more than one BRAM Partition will be needed after the register levels. However, instead of replicating the tree again as the previous implementation we take another approach. In this approach we split the tree both horizontally and vertically rather than splitting only horizontally as the prior implementations. Horizontal partitioning splits the tree level by level, vertical partitioning splits the tree as left and right subtrees. In this implementation which can be seen from Fig. 3, the last level of the registers are counted as the root nodes of its left and right subtrees which are stored inside BRAM

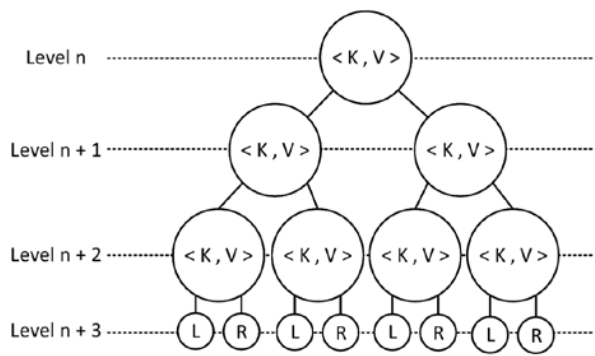

Figure 3. Connection of Register and Partition Layers

Partitions. When a key is searched and can not be found inside the last level of the registers and if the key is bigger than the last register it was compared, it continues to be searched in its right BRAM Partitions. Otherwise it is searched in the left partitions. The number of the subtrees and the number of nodes stored in the registers are configurable.

With this functionality if the keys on the last register level are found to be directed to other subtrees, unlike the horizontal partitioning, more than 2 keys (1 for single port BRAMs) can be searched in parallel.

### 3) Addition of Buffers

The previous implementation does not totally remove the bottleneck. In worst case, more than two keys can be directed to be searched inside the same BRAM Partition resulting in a stalling process. Therefore we add a buffer for each of the subtrees as can be seen from Fig. 4. These buffers' slot number is configurable and the buffers are holding keys that can not be searched in that clock cycle due to the port limit. The next cycle the keys in the buffer will be fetched and directed to their respected partitions.

After the key search on the last level of the register layer, the subtree that the key needs to be searched in is found. However, since the number of keys that are going to be transferred to the same subtree are not known, it is unknown to which index of the buffer the key should be put. Therefore a parallel technique to find which slot is suitable for the key needs to be developed.

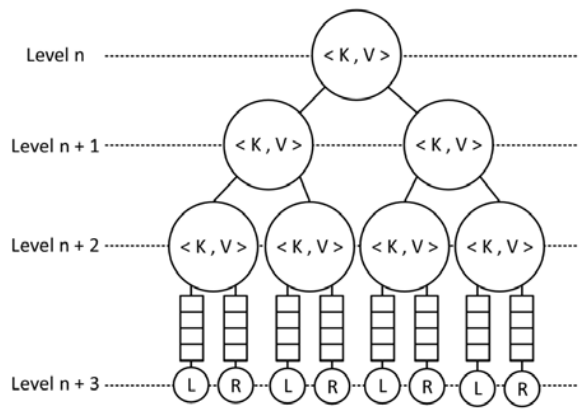

Figure 4. Addition of Buffers To Decrease Stalling Rates

To resolve this issue we implemented two different mapping approaches for the buffers.

1. Direct mapping
2. Queue mapping

Direct mapping approach directly puts the key to the buffer's slot which is equal to the key's index in the chunk. That is, if the fifth key is going to be routed to the sixth subtree, the key is put to the sixth subtree's buffer's fifth slot. While using dual ports, buffers are divided to two, first half holds the keys that are going to be searched with the first port, the second half holds the keys to be searched with the second port. One key is selected for the first port and another key is selected for the second port.

If a key's mapped slot is already full then the register layers of the tree stalls, prohibiting any new keys to enter the search process. The downside of this approach is, even though the buffer's other slots are empty, stalling can occur due to mapped slot's unavailability. An example is given in Fig. 5, showing the relevant keys, which should be directed to the related BRAM Partition, in the incoming key list stored in a buffer with no conflicts shown in subfigure (a) and stored in a buffer which already has keys inside that creates conflict due to the need of storage in an already occupied slot, shown in (b). The search in the second subfigure gets stalled in order to free up the occupied slots, whereas there is no stalling in the first subfigure. For the sake of simplicity, we do not add a stalling key selection in buffers, meaning that the key which creates the stall is not necessarily selected but the key which comes earlier in the buffer is selected.

In the first approach the search can stall even though there are free buffer slots therefore we implemented the Queue mapping approach in order to decrease the stalling rates. Along with the buffers, we are storing a list of pointers which contains read and write pointers to be used by the buffers. Each buffer has a unique read and write pointer. Read pointer shows the index of the first key that will be fetched from the buffer and transferred to the partition layers. Write pointer shows the available index that the initial next key can be stored. These pointers can be thought as the beginning and ending markers of the queue.

At the end of the register layers, we label the remaining keys with an id of the subtree that they are going to be transferred. Therefore the next clock cycle the number of keys that are going to be put to the same buffer is calculated and given a number. For instance if two keys will be in the same buffer, first key gets labeled with 0 and the second gets labeled with 1.

After the labeling on the indexes for storage in buffers, since every key has a label showing how many keys are going to be stored before it and the keys are put to the slot which is shown by the index of the write pointer of that buffer added by the label index of the key. The keys are started to be fetched from the index where the read pointer shows. With the usage of pointers the keys that are searched preserve their orders that is the first key that is added to the search is processed first.

If a buffer is no longer capable of acquiring a new key due to being full, that is if the sum of the key's index and the buffer's write pointer's value pointing a non-empty buffer slot, then the register layers of the tree stalls, prohibiting any new keys to enter the search process. However the partition layers continue on producing results, and fetching the keys from buffers, decreasing the number of keys remained inside the buffers. As soon as all of the buffers have at least one empty slot, the stalling process finishes and new keys are let to enter the search process. The behaviour of the Queue mapping approach is shown in Fig. 6, which is the same scenario used in the prior figure. However with this approach there are no conflicts happening since the buffer is not full and the newly wanted keys are put inside of the buffer without creating stalls. The conflict only appears if there are less empty slots than the number of keys that needs to go to the same buffer.

III. RESULTS AND DISCUSSION

We implement the discussed methods via Bluespec SystemVerilog (BSV) on VC709 platform, with Virtex7 FPGA. This platform has 1470 BRAMs with the size of 36 kbits each.

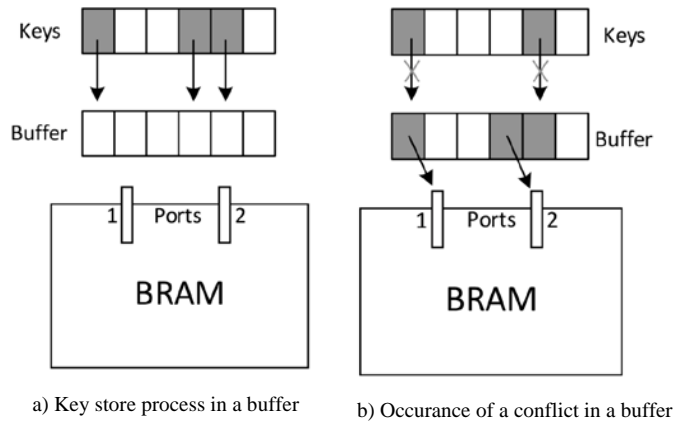

a) Key store process in a buffer    b) Occurance of a conflict in a buffer

Figure 5. Behaviour of buffers in Direct Mapped Implementation

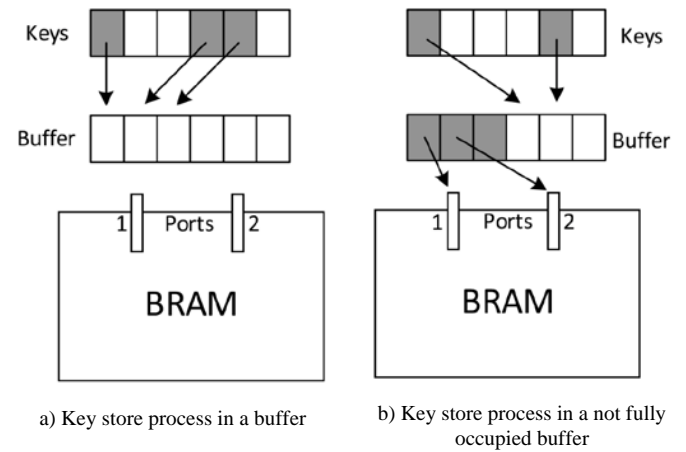

a) Key store process in a buffer    b) Key store process in a not fully occupied buffer

Figure 6. Behaviour of buffers in Queue Mapped Implementation

For different configurations, we run different sets of keys on the implementations of:

- Horizontal Partitioned (tagged as Hrz)
- Duplicated Horizontal Partitioned - 4 Trees (Dup4)
- Duplicated Horizontal Partitioned - 8 Trees (Dup8)
- Hybrid Partitioned - Direct Mapping, 4 Trees (Hyb4)
- Hybrid Partitioned - Queue Mapping, 4 Trees (Hyb4q)
- Hybrid Partitioned - Direct Mapping, 8 Trees (Hyb8)
- Hybrid Partitioned - Queue Mapping, 8 Trees (Hyb8q)

Hrz is similar to the implementation shown in [1] and considered as the baseline implementation. Although Hybrid Partitioned implementations are selected as 4 or 8 trees, the tree numbers were selected only for simplicity and the numbers can be configured in a fully flexible manner at compile time. It is important to clarify that, the tree numbers show the number of subtrees of the single tree for the hybrid implementations, they are not the number of duplications, which is the case for duplicated implementations. We also tuned the implementations as Addition of Buffers section to make the comparisons more just, that is we put the tree's first levels to registers.

To have a comprehensive evaluation, we repeat the experiments with three different key sets. The key sets that we run consists of 64K and 256K versions of:

- Equal, same key which is selected as a leaf node, worst case scenario.
- Random, keys are selected randomly.
- Split, keys are from different subtrees, best case scenario

With the Equal key set since all of the searched keys are equal, they would all need to go through the same path to be found creating conflicts and stalls. Random keys will have random paths. Split key set are consisted of keys which are from different subtrees and they will pass through different paths therefore they will not create any conflicts.

We run our implementations using different sizes of the key sets that were explained priorly and we recorded the required number of cycles for all of the keys to be found. Fig. 7 shows the acceleration rate of the search done, relative to the baseline Horizontal implementation (Hrz), with the key sets of 64K and 256K sizes. The speed up for the different sizes of the key sets do not show distinguishable differences, due to the fact that the implementations still produce the same or close throughput. For instance, a Dup4 implementation will always have a 4 times more throughput regards of Hrz regardless of the key size because it has 4 times more ports that can conduct a search.

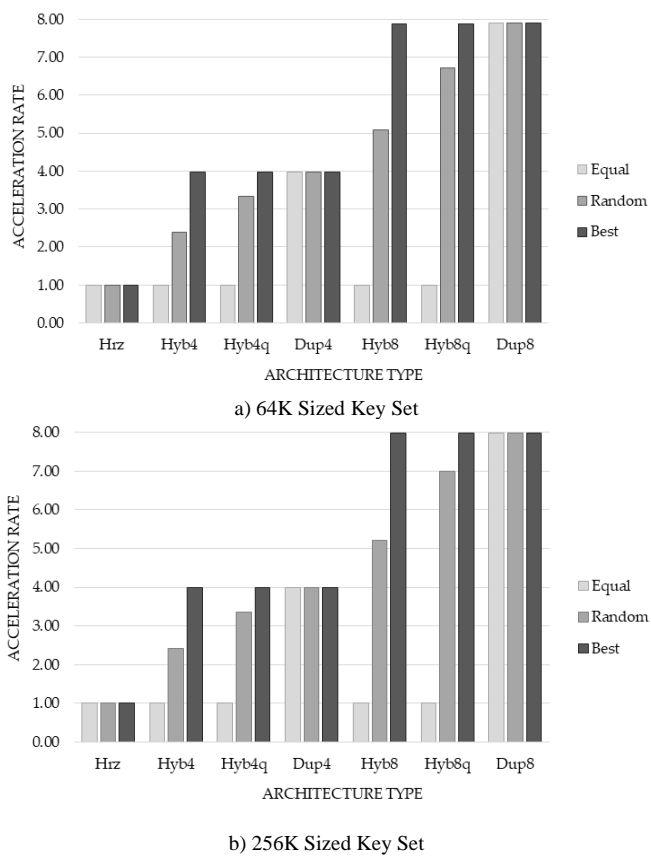

a) 64K Sized Key Set

b) 256K Sized Key Set

Figure 7. Horizontal Relative Acceleration Rate Results for Different Sized Key Sets

As can be seen from the figure since Dup8 implementation has a total of 8 Trees, resulting to 8x2 ports, that can be searched at the same time, it has the least amount of cycles needed in order to get the results from the key sets. Dup8 and Dup4 implementations have a constant rate relative to Horizontal implementation, the reason for this is since they do not stall the types of keys does not matter. However the Hybrid implementations stall with the Equal key sets and since they only have 1 Tree, the number of ports is equal to the Horizontal implementation so they converge to Horizontal's results. With random key sets, Hybrid implementations stall but in a lesser rate than the previous one and split key set does not create a stall since all of the keys are separated in different subtrees. The difference of the direct mapped and queue based implementations occurs due to the direct mapped version's more increased stall rate than queue based version's.

It can be seen that Dup8 version has nearly a 16 key/cycle throughput and the key/cycle throughput rate increases proportionally with the number of trees for all implementations. Since the implementations can be configured with more trees better throughput can be acquired however the bandwith of the memory system should be taken into consideration while determining the number of trees.

The Horizontal and Duplicated implementations can be considered as worst and best baselines for other implementations since they do not create any stalls. Although the results show Dup8 as the most desirable system, it must not be forgotten that Dup8 needs nearly 8 times more memory space than Horizontal and Hybrid implementations. For a dataset that contains $2^{20}$ nodes, Dup8 contains $2^{23}$ nodes, therefore these implementations decrease the size of the dataset, can be stored inside the BRAMs, that we can conduct a search on. With a bigger dataset, as long as it fits to FPGA memory, the throughputs will stay similar but the latencies will show differences because it will take more time to reach the leaf nodes.

As can be seen from Fig. 8, memory usage of Duplicate implementations recorded higher than the other implementations due to the need of duplication of the same data. In the hybrid implementations the memory usage is lower however due to the buffer system they are in need of more slice LUTs.

The queue implementation takes more space than the implementation with direct mapping, since it needs to hold both write and read pointers for each of the buffers and the routing for the keys to the buffer slots can change therefore it requires additional hardware. Direct mapping implementation takes less space due to its simplicity for the routing of the keys to the buffers. Although it needs less hardware to function, it has an increased chance to stall compared to the first implementation since the new keys will not be put to the empty slots directly and there can be a stalling even though there are empty slots. Therefore it is better to use the direct mapping implementation when the hardware resources are limited and the keys are diverse, that is the keys would not be going to the same subtree often.

Fig. 9 shows that due to the additional hardware needs of hybrid implementations they have a higher energy consumption and they need a slower frequency to function, especially queue implementations. The slower frequency need occurs because of the fact that in the queue based implementation prior to putting the keys to the buffers, we are checking which key needs to go to which slot and this is done by checking the keys one by one resulting a longer critical path. This critical path can be shortened by pipelining the checking algorithm however because of the incoming new key chunks, either we need to stall or need more storage space not to lose any key chunks. In our implementations the size of the buffers are equal to the number of the keys that can be searched at most, that is Dup4, Dup4q has buffer sizes of 8 whereas Dup8, Dup8q has buffer sizes of 16. The buffer sizes are fully configurable and can be increased in order to achieve less stall rates, however the trade-off between throughput and memory space should be taken into consideration.

In our implementation results, between the hybrid implementations with the configurations that are mentioned prior, we can see that there is an acceleration difference of 32% to 39% for the random set in favor of the queue implementations, and a difference of 7% to 8% in terms of clock frequency in favor of the direct mapped implementations.

## IV. RELATED WORK

Hardware acceleration using GPUs, FPGAs, and Application-Specific Integrated Circuits (ASICs) is a promising approach to achieve higher throughput and lower energy in comparison to conventional CPUs. It has been shown that many state-of-the-art applications such as neural networks [16, 17], query processing [17-19], and autonomous cars [20] can take the advantage of hardware-based acceleration. Among

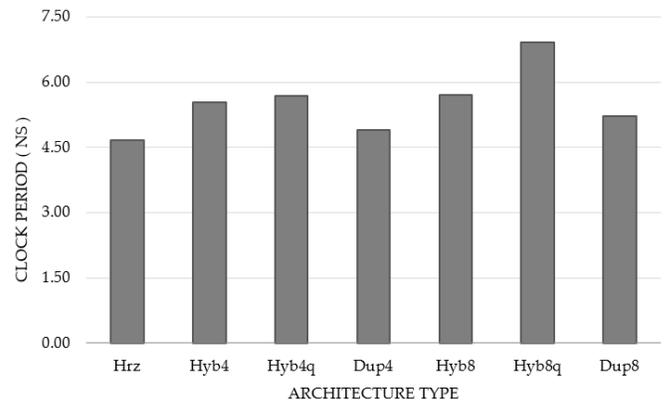

a) Clock Period Results ( ns )

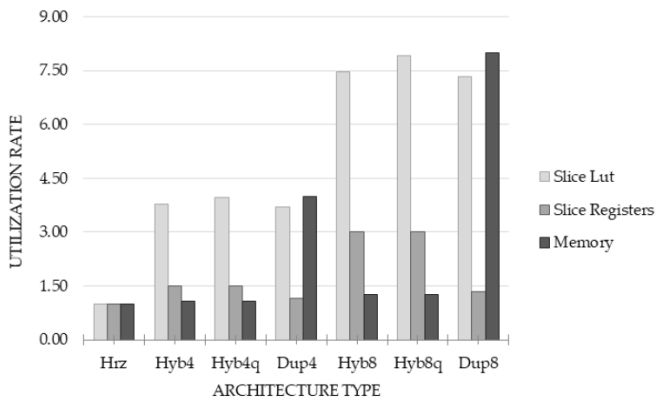

Figure 8.  Horizontal Relative Utilization Results

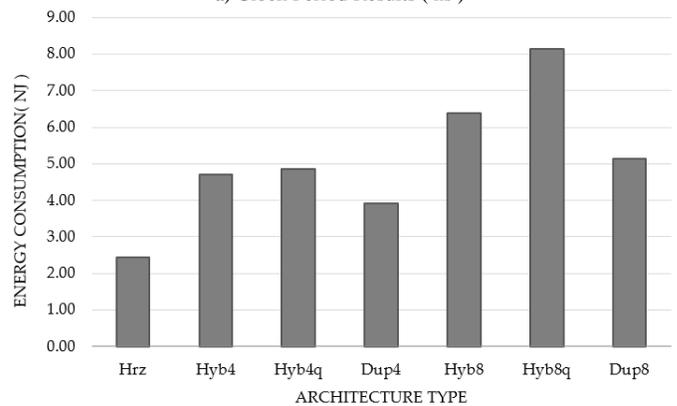

b) Energy Consumption Results ( nJ )

Figure 9.  Timing and Energy Results

these approaches, FPGA-based designs have unique characteristics such as more flexibility versus ASICs and more optimized compared to GPUs, which is making them increasingly more popular.

BST searching and construction has also been considered for the acceleration using FPGAs, due to its high-demand for high-throughput and low-energy. For instance, [1] presents an FPGA-based BST construction and searching algorithm; however, the throughput achieved is not the maximum possible, as only horizontal tree partitioning is considered. In the same line, there are other efforts [2-5], in which mostly the pipeline capabilities of FPGAs have been considered as the optimization point. Also, recently a multi-FPGA version of this algorithm has been presented [6]. In comparison to mentioned works above, the accelerator presented in this paper more efficiently takes the advantage of FPGA architecture by a hybrid horizontal and vertical tree partitioning among the FPGA BRAMs. Our approach provides relatively higher-throughput, thanks to the effective exploitation of inherent parallelism combined with the dataflow execution model.

## V. CONCLUSION

In order to achieve our goal i.e., to create a deeply pipelined and massively parallel binary tree search accelerator leveraging FPGAs, we presented the tree duplication, and horizontal-vertical partitioning of tree levels into the on-chip memories. To reduce the bottleneck created by the number of ports available, we added buffers to decrease the stall rates.

Duplicated tree implementations have the highest throughput ratio however since they need to create copy of the same tree they need more memory space. The hybrid implementations are not creating copies of the dataset therefore they need less memory space than their counterparts. However they are prone to stalling and the direct mapping has a high chance for stalling. The direct mapping is less complex than the queue based hybrid implementation therefore resulting to a need for less hardware.

If high memory usage would not create any problems, the duplicate approaches will have the greatest throughput. If high memory usage will create problems than the domain of the keys that will be searched should be taken into consideration. If the keys will have a higher stall rate the queue based solution should be selected, otherwise direct mapped implementation should be selected.

We are working on the extension of this work to cover the BST construction phase by adding Delete and Insert operations. Also, we will extend the techniques proposed in this paper to support big data sets by exploiting off-chip storages as well exploiting energy-efficiency techniques [21, 22] such as voltage underscaling [23-25].

## ACKNOWLEDGMENT

The research leading to these results has received funding from the European Union's Horizon 2020 Program under the LEGaTO Project (www.legato-project.eu), grant agreement n° 780681.